\documentclass[ twocolumn]{aastex631}
\usepackage{comment}
\usepackage[version=4]{mhchem}

\usepackage[english]{babel}
\usepackage{graphicx}
\usepackage{natbib}

\begin{document}

\title{Synthesis and Spectroscopic Characterization of Interstellar Candidate Ethynyl Thiocyanate: HCCSCN}


\author[0000-0001-5816-4102]{Elena R. Alonso}
\affiliation{Grupo de Espectroscopía Molecular (GEM),  \\
Edificio Quifima, Laboratorios de Espectroscopia y Bioespectroscopia, \\
Universidad de Valladolid, 47011 Valladolid, Spain}

\author[0000-0002-9291-1762]{Aran Insausti}
\affiliation{Departamento de Química Física, Facultad de Ciencia y Tecnología,\\ Universidad del País Vasco (UPV/ EHU), 48940 Leioa, Spain.}

\author[0000-0003-3567-0349]{Lucie Kolesniková}
\affiliation{Department of Analytical Chemistry, University of Chemistry and Technology,\\ Technická 5, 16628 Prague 6, Czech Republic}

\author[0000-0002-1992-935X]{Iker León}
\affiliation{Grupo de Espectroscopía Molecular (GEM),  \\
Edificio Quifima, Laboratorios de Espectroscopia y Bioespectroscopia, \\
Universidad de Valladolid, 47011 Valladolid, Spain}

\author[0000-0003-1254-4817]{Brett A. McGuire}
\affiliation{Department of Chemistry, Massachusetts Institute of Technology, Cambridge, MA 02139, USA}
\affiliation{National Radio Astronomy Observatory, Charlottesville, VA 22903, USA}

\author[0000-0002-5171-7568]{Christopher N. Shingledecker}
\affiliation{Department of Chemistry, Virginia Military Institute, Lexington, Virginia 24450, United States}
\author[0000-0003-3248-3564]{Marcelino Agúndez}
\affiliation{Instituto de F\'isica Fundamental, CSIC, Calle Serrano 123, 28006 Madrid, Spain}

\author[0000-0002-3518-2524]{José Cernicharo}
\affiliation{Instituto de F\'isica Fundamental, CSIC, Calle Serrano 123, 28006 Madrid, Spain}
\author[0000-0002-2887-5859]{Víctor M. Rivilla}
\affiliation{Centro de Astrobiolog\'ia (CSIC-INTA), Ctra. de Ajalvir Km. 4, Torrej\'on de Ardoz, 28850 Madrid, Spain}

\author[0000-0002-1254-7738]{Carlos Cabezas}
\affiliation{Instituto de F\'isica Fundamental, CSIC, Calle Serrano 123, 28006 Madrid, Spain}

\author[0000-0003-4493-8714]{Izaskun Jim\'enez-Serra}
\affiliation{Centro de Astrobiolog\'ia (CSIC-INTA), Ctra. de Ajalvir Km. 4, Torrej\'on de Ardoz, 28850 Madrid, Spain}

\author[0000-0003-4561-3508]{Jesús Mart\'{\i}n-Pintado}
\affiliation{Centro de Astrobiolog\'ia (CSIC-INTA), Ctra. de Ajalvir Km. 4, Torrej\'on de Ardoz, 28850 Madrid, Spain}

\author[0000-0002-2929-057X]{Jean-Claude Guillemin}
\affiliation{Univ Rennes, Ecole Nationale Supérieure de Chimie de Rennes, CNRS, ISCR- UMR 6226,Rennes, France}

\begin{abstract}

This work aims to spectroscopically characterize and provide for the first time direct experimental frequencies of the ground vibrational state and two excited states of the simplest alkynyl thiocyanate (HCCSCN) for astrophysical use. Both microwave (8-16~GHz) and millimeter wave regions (50-120~GHz) of the spectrum have been measured and analyzed in terms of Watson's semirigid rotor Hamiltonian. A total of 314 transitions were assigned to the ground state of HCCSCN and a first set of spectroscopic constants have been accurately determined. Spectral features of the molecule were then searched for in Sgr B2(N), NGC 6334I, G+0.693-0.027 and TMC-1 molecular clouds. Upper limits to the column density are provided.

\end{abstract}

\keywords{Catalogs (205) --- Molecular spectroscopy (2095) --- Interstellar molecules (849) --- Laboratory astrophysics (2004) ---  Experimental data (2371)}

\section{Introduction} \label{sec:intro}

In the last years, there has been a renewed interest on the chemistry of sulfur in the interstellar medium, aiming to shed light on the long-standing problem of the missing sulfur. That is, in cold dense clouds, all detected S-bearing molecules only account for a very small fraction of sulfur, meaning that the main reservoir of sulfur is unknown \citep{vidal_reservoir_2017}. Motivated by this interest, a large variety of sulfur-bearing molecules have been detected in the interstellar medium on the basis of their pure rotational spectrum. Recent detections of S-bearing molecules include S$_2$H \citep{Fuente2017}, NS$^+$ \citep{cernicharo2018}, HCS, HSC \citep{Agundez2018}, HC$_3$S$^+$ \citep{Cernicharo2021a}, NCS, HCCS, H$_2$CCS, H$_2$CCCS, C$_4$S, C$_5$S \citep{Cernicharo2021b}, HC(O)SH, C$_2$H$_5$SH \citep{Rodriguez-Almeida2021}, HCSCN, HCSCCH \citep{Cernicharo2021c}, HCCS$^+$ \citep{Cabezas2022}, HC$_4$S \citep{Fuentetaja2022}, HSO \citep{Marcelino2023}, HCNS \citep{Cernicharo2024}, HOCS$^+$ \citep{Sanz-Novo2024a}, and HNSO \citep{Sanz-Novo2024b}. Although the problem of the missing sulfur remains open, this plethora of detections has shown that sulfur participates in a particularly rich chemistry. 

A particularly interesting case of S-bearing molecules is the CHNS family, in which three different isomers have been detected in the interstellar medium. The three detected isomers, in increasing energy, are thioisocyanic acid (HNCS, \citealt{Frerking1979}), thiocyanic acid (HSCN, \citealt{Brunken2009a,Halfen2009,Adande2010}), and thiofulminic acid (HCNS, \citealt{McGuire2016HNCS,Cernicharo2024}). The most stable isomer is HNCS, while HSCN and HCNS lie 3200 K and 17300 K above, respectively \citep{Wierzejewska2003,McGuire2016HNCS}. In spite of this ordering in energy, it turns out that in the cold starless core TMC-1, the most abundant isomer is HSCN, which is not the most stable one \citep{Cernicharo2024}. Given that these molecules are relatively abundant in interstellar clouds such as Sgr B2N, NGC 63341, G+0.693$-$0.027, TMC-1, derivatives in which a CN or CCH group substitutes one hydrogen atom are good targets for astronomical detection. Taking into account the observed abundance of HSCN, it seems promising to search for the CN and CCH derivatives of HSCN, in particular HCCSCN 
which is an isomer of the recently detected NCCHCS species in TMC-1 \citep{Cabezas_2024}.

\begin{figure}[ht]
   \centering
   \includegraphics[trim = 0mm 0mm 0mm 0mm, clip, width=8.2cm]{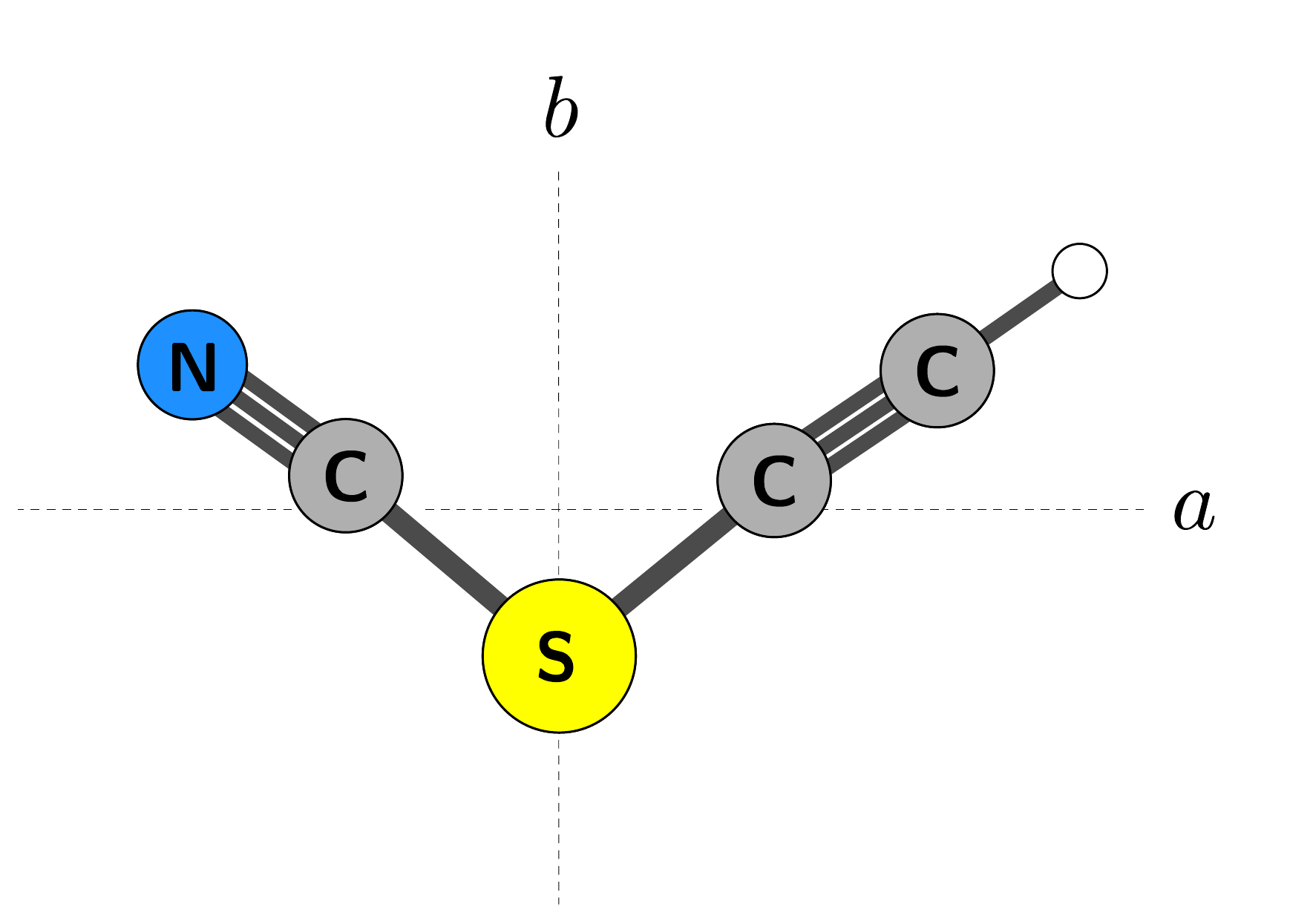}
      \caption{Optimized structure of the HCCSCN molecule depicted in the $ab$ inertial plane.
      }
      \label{molecule}
   \end{figure}

In this paper, we present the first measurements of the pure rotational spectrum of ethynyl thiocyanate (HCCSCN), the CCH derivative of HSCN (see Figure\,\ref{molecule}), and an astronomical search for it toward several molecular clouds. The molecule has been first synthesized by \cite{Lu2021} and only characterized by UV-Vis and IR spectroscopy. HCCSCN is predicted as the third most stable acyclic structure in the HC$_{3}$NS family of isomers, right after NCC(H)CS and HCCNCS \citep{Lu2021}.

\section{EXPERIMENTAL METHODOLOGY} \label{sec:experimental}
Searching for and detecting a new molecule in the ISM typically begins on Earth at a spectroscopic laboratory. This work has been carried out in synergy between an organic chemist, spectroscopists and astrophysicists. This work tandem has been very successful in the detection of more than 310 molecules in the interstellar medium or circumstellar shells up to the present day \citep{endres_cologne_2016}. The followed working procedure to obtain a set of parameters that the astrophysicist uses to carry out the complex search for molecules in the different regions of the ISM, includes, in many cases, complex chemical synthesis, the use of home-built and expensive experimental techniques, as well as computational methodology for the calculation of molecular structures.

\subsection{Chemical Synthesis} \label{subsec:synthesis}

The ethynyl thiocyanate has been synthesized by reacting alkynyl(phenyl)iodonium trifluoromethanesulfonate with sodium thiocyanate \citep{Lu2021}. In this work, it has been prepared by reacting tributylethynylstannane with sulfur cyanide, with a yield $63\%$. A similar approach has been used in the past to synthesize 1,2-propadien-1-ylthiocyanate. The compound, which is kinetically unstable, was stabilized in a high-boiling solvent (diethyleneglycol dibutyl ether) in the presence of small amounts of duroquinone (a high-boiling radical inhibitor), and conserved in dry ice.

      \begin{figure*}[ht]
   \centering
   \includegraphics[width=18.0cm]{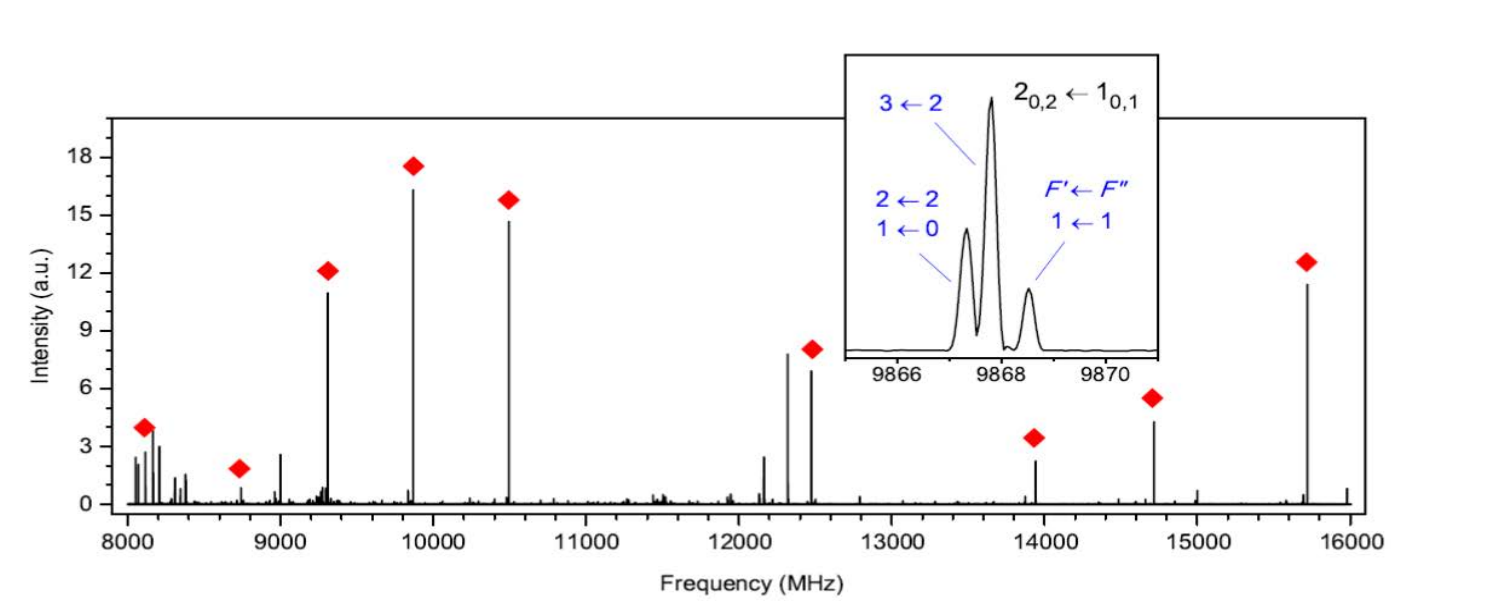}
      \caption{Broadband CP-FTMW spectrum of HCCSCN. \textbf{The red rhombus indicates measured transition}. The inset shows the nuclear quadrupole hyperfine structure for $2_{0,2}\leftarrow 1_{0,1}$ rotational transition.
      }
      \label{CP}
   \end{figure*}

\subsection{Rotational spectra} \label{subsec:spectra}

In the first step, the broadband rotational spectrum of HCCSCN was recorded from 8-21~GHz employing the Chirped-Pulse Fourier Transform Microwave (CP-FTMW) spectrometer of the Universidad de Valladolid \citep{CP-FTMWVALLADOLID,Leon2019}. The synthesized liquid sample, highly volatile, was deposited in a small sealed container through which a carrier gas, Neon, was passed at a pressure of 10-12~bar. This container was introduced in a cooling bath with dry ice to minimize the quick decomposition of the sample. The molecules seeded in the carrier gas were adiabatically expanded through a nozzle (1~mm in diameter) and probed in the cold and efficiently isolated conditions of a supersonic expansion. Then, sequentially, the molecules were polarized using a high power excitation pulse of 300~W, to then capture up to 60000 free induction decays that were averaged in the time domain and Fourier transformed to obtain the spectrum in the frequency domain. The Figure \ref{CP} shows the obtained CP-FTMW spectrum.

After completing the broadband microwave range experiment, we moved on to the millimeter-wave range, recording the room temperature rotational spectrum using the Valladolid home-built millimeter wave absorption spectrometer \citep{Daly2014,Kolesnikova2017a}. Employing a multiplication (4x and 6x, VDI. Inc) of the basic synthesizer frequency (Agilent, 250~kHz-20~GHz), the frequency range from 50-120~GHz was reached. The synthesizer output was frequency modulated at 10.2~kHz \textbf{with a} modulation depth between 50 and 60~kHz. A Schottky diode detector (VDI, Inc.) was used to capture the signal after a double radiation pass through the cell, and then it was introduced in a lock-in amplifier. A second derivative shape of the lines resulting from 2f detection was fitted to the Gaussian profile function. The uncertainty of the line center frequency is estimated to be better than 50~kHz.

\subsection{Quantum-chemical calculations} \label{subsec:quantum}

To be able to interpret the rotational spectra, it is very useful to have theoretically predicted rotational parameters. For this, it is necessary to have an optimized geometry of the molecule, which can be obtained using computational calculations. Since HCCSCN has only 6 atoms, a triple bond between the carbons and a thiocyanate group, it can form only a single conformation, thereby simplifying the computational process.

The geometry optimization was done employing both DFT and \textit{ab initio} computational methods implemented in Gaussian 16 \citep{g16}. The levels of theory used were the B3LYP density functional \citep{Becke1993,Becke1992}, including the Grimme D3 dispersion interactions \citep{Grimme2010}  with Becke–Johnson damping \citep{Grimme2011}), and the Møller–Plesset second-order method \citep{MP2}, respectively. The basis set employed in both cases is the Pople split-valence triple-zeta basis set augmented with diffuse and polarization functions on all atoms (the 6-311++G (d,p) basis set \citep{Frisch1984}).

Furthermore, the frequencies of the vibrational modes were also calculated at the same B3LYP density functional and the same basis set employed in the geometry optimization mentioned above. The calculation of the first-order vibration-rotation correction constants $\alpha_{i}$, that define the well-known vibrational dependence of rotational constants, was used to derive the changes in rotational constants according to the equation: $B_{v}=B_{e}-\sum_{i}\alpha_{i}(v_{i}+1/2)$, where $B_{v}$ and $B_{e}$ substitute all three rotational constants in a given vibrational state and in equilibrium, respectively. $v_{i}$ is the vibrational quantum number of the \textit{i}th vibrational mode. This allowed to assign the two excited states found in our spectra based on the differences between theoretical and experimental rotational constants.

\section{Analysis of the spectra} \label{sec:analysis}

The observation of characteristic transition patterns is extremely helpful in searches and assignments of the spectrum of a new species and, if possible, it is useful to start at the lower frequency range. We therefore first relied on our broadband spectra measurements carried out with the CP-FTMW spectrometer. HCCSCN is a prolate asymmetric rotor and is predicted to have dipole moment components along the $a$ and  $b$ inertial axes. ($|\mu_{a}|=$~3.2~D, $|\mu_{b}| =$~1.2~D).



The recorded broadband spectrum indeed presented very intense lines. The assignment of these intense lines was straightforward to the $a$-type R-branch typical triad of transitions: $2_{0,2}\leftarrow1_{0,1}; 2_{1,1} \leftarrow 1_{1,0}; 2_{1,2}\leftarrow1_{1,1}$ and $3_{0,3}\leftarrow2_{0,2}; 3_{1,3}\leftarrow2_{1,2}; 3_{1,2}\leftarrow2_{1,1}$. Then $b$-type R-branch transitions were also measured. We could observe that all these transitions present hyperfine structure (see Figure \ref{CP}) due to the presence of $^{14}$N in the molecule, with nuclear spin (\textit{I} = 1) and nonzero electric quadrupole moment, which interacts with the electric field gradient of the molecule at the position of this nucleus and causes the splitting of each rotational level into several sublevels \citep{Gordy}. A total of 13 transitions with 43 hyperfine components were measured with an assumed frequency accuracy of 10~kHz. A first set of rotational constants was obtained using a Watson's A-reduced semi-rigid rotor analysis in the I$^{r}$-representation, adding a term to account for the nuclear quadrupole coupling contribution \citep{Foley1947,Robinson1953}, enabling the determination of the $^{14}$N nuclear quadrupole coupling constants. \textbf{Some lines in the spectrum remain unassigned and could be attributed to transitions of some compounds from the synthesis or some perturbed and unassigned excited vibrational states. The most prominent spectral features are accounted in our analyses and follow the expected spectral pattern of HCCSCN.}

      \begin{figure}[t]
   \centering
   \includegraphics[trim = 5mm 50mm 30mm 5mm, clip, width=9.0cm]{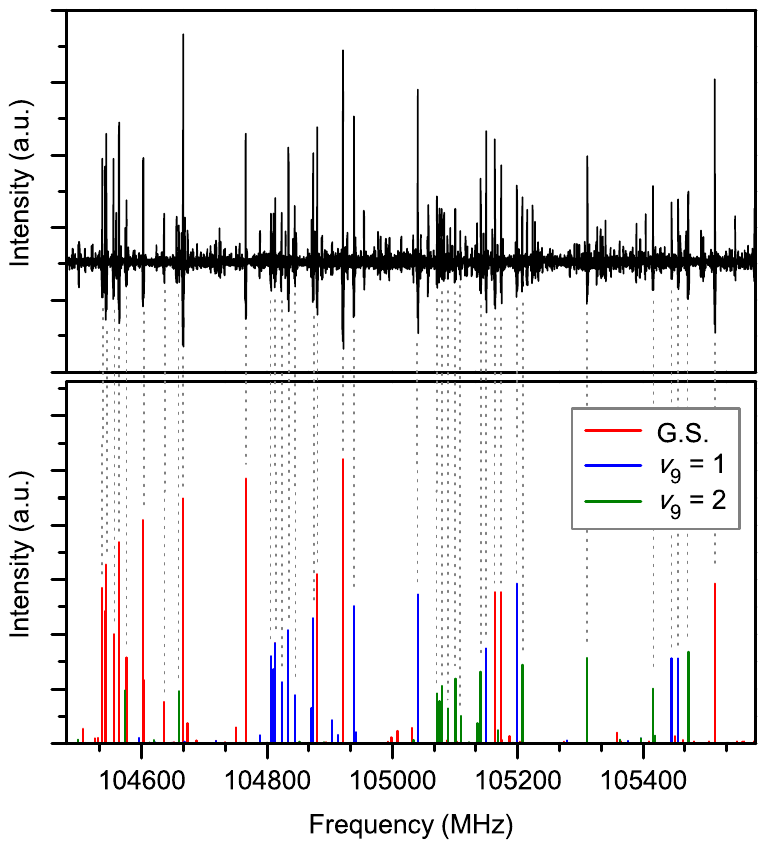}
      \caption{Section of the room-temperature millimeter wave spectrum of HCCSCN. The upper panel shows the experimental spectrum while the lower panel illustrates a stick reproduction of the ground state (G.S.) and $v_{9}=1,2$ excited vibrational states using the spectroscopic constants from Table \ref{constants}. \textbf{The trend of the relative intensities is not always consistent in theory and experiment due to the performance of the experimental set-up (multiplication chain, cell transmision and detector sensitivity)}}
      \label{mw}
   \end{figure}

\begin{table*}
\caption{Spectroscopic constants of HCCSCN in the ground state (G.S.) and two excited vibrational states ($A$-reduction, I$^{\text{r}}$-representation)
in comparison with quantum-chemical calculations.}
\label{constants}
\setlength{\tabcolsep}{12.0pt}
\begin{center}
\begin{tabular}{ l r r r r r }
\hline\hline
Parameter &  G.S.$^{a}$ & $v_{9}=1$$^{a}$ & $v_{9}=2$$^{a}$ & Calculated$^{b}$ \\
\hline
$A                                      $  /MHz             &  10294.7535(31)   &  10390.088(34)   &  10486.759(48)   &  10458      \\
$B                                      $  /MHz             &   2771.73159(44)  &   2782.19440(92)  &   2792.5389(18)  &   2717     \\
$C                                      $  /MHz             &   2179.00087(41)  &   2180.95866(55)  &   2182.85772(97)  &   2156     \\
$\mathit{\Delta_{J}}                    $  /kHz             &      1.95771(80)  &      1.97331(69)  &      1.9898(13)  &      1.78     \\
$\mathit{\Delta_{JK}}                   $  /kHz             &    -32.1930(30)   &    -32.4694(59)   &    -32.759(11)   &    -31.02      \\
$\mathit{\Delta_{K}}                    $  /kHz             &    195.19(63)     &    201.7(13)     &    211.1(19)     &    199.07        \\
$\delta_{J}                             $  /kHz             &      0.76512(25)  &      0.77350(38)  &      0.78425(75)  &      0.68     \\
$\delta_{K}                             $  /kHz             &      2.843(21)    &      3.570(30)    &      4.365(47)    &      2.39       \\
$\mathit{\Phi_{J}}                      $  /mHz             &       9.11(78)    &         9.11$^{g}$     &          9.11$^{g}$   &     8.64                 \\
$\mathit{\Phi_{JK}}                     $  /mHz             &       -87.5(37)     &     -95.6(61)  &      -87.5$^{g}$   &    -64.39    \\
$\mathit{\Phi_{KJ}}                     $  / Hz             &       -1.1064(59)  &   -1.1064$^{g}$   &        -1.1064$^{g}$   &   -1.21   \\
$\mathit{\Phi_{K}}                      $  / Hz             &     [11.85]$^{f}$   &   [11.85]$^{f}$        &    [11.85]$^{f}$      &     11.85   \\
$\phi_{J}                               $  /mHz             &       [ 4.14]$^{f}$    &  [ 4.14]$^{f}$    &    [ 4.14]$^{f}$      &   4.14     \\
$\phi_{JK}                              $  /mHz             &    [ 5.47]$^{f}$  &   [ 5.47]$^{f}$         &     [ 5.47]$^{f}$  &    5.47   \\
$\phi_{K}                               $  / Hz             &   [1.36]$^{f}$  &   [1.36]$^{f}$ &      [1.36]$^{f}$ &   1.36  \\
$\chi_{aa}                            $  /MHz             &    -1.5260(73)               &      -             &         -          &      -1.69                \\
$\chi_{bb}                            $  /MHz             &         0.290(10)          &      -             &       -            &        0.49              \\
$\chi_{cc}                            $  /MHz             &     1.236(10)              &     -              &       -            &       1.20               \\
$J_{\text{min}}/J_{\text{max}}$         &     1/25        &      9/26             &        11/24           &                   &                      \\
$K_{a}^{\text{min}}/K_{a}^{\text{max}}$ &     0/20        &      0/15             &        0/7           &                   &                      \\
$N$$^{c}$                               &     314           &      165          &    43            &                   &                      \\
$\sigma_{\text{fit}}$$^{d}$/MHz         &     0.020         &     0.043              &     0.028              &                   &                      \\
$\sigma_{\text{w}}$$^{e}$               &     0.88          &      0.87             &       0.94            &                   &                      \\
\hline
\end{tabular}
\end{center}
\tablecomments{
$^{a}$The numbers in parentheses for all the experimentally determined parameters are their uncertainties in units of the last decimal digits. Their values are close to 1$\sigma$ standard uncertainties because the unitless (weighted) deviation of the fit is close to 1.0. SPFIT/SPCAT program package \citep{Pickett1991} was used for the analysis.
$^{b}$Equilibrium values calculated at B3LYP/6-311++G(d,p) level of theory.
$^{c}$Number of distinct frequency lines in the fit. All measured transitions are available in a machine-readable form in the online journal.
$^{d}$Root mean square deviation of the fit.
$^{e}$Unitless (weighted) deviation of the fit.
$^{f}$Values in brackets fixed to those calculated theoretically.
$^{g}$Values fixed to the ones fitted for G.S}.

\end{table*}


With this first set of rotational parameters a prediction of the transitions at higher frequencies was done, and we proceeded to the analysis of the millimeter-wave spectrum in the 50-120~GHz frequency range.  The prediction presented a shift in the frequency of the transitions with respect to the experimental ones due to the centrifugal distortion effects, but it was not difficult to find the transition patterns that led to measure a total of 271 additional transitions of the molecule with an assumed frequency accuracy of 30~kHz. In this frequency range, the hyperfine structure collapses into a single line, simplifying the spectra. The final fit, combining the transitions from both microwave and millimeter-wave experiments with their corresponding weights, was obtained using Watson’s A-reduced Hamiltonian in the I$^r$-representation \citep{Watson1977} and the 
 Pickett’s SPFIT/SPCAT program suite \citep{Pickett1991}. (All measured transitions are available in a machine-readable form in the online journal.) The spectroscopic data obtained in this work allowed the determination of the rotational constants, nuclear quadrupole coupling constants as well as the full set of quartic and some sextic centrifugal distortion constants presented in Table \ref{constants}. This spectroscopic information thus enables a search for this newly characterized molecule in different regions of the ISM.

Further inspection of the spectrum revealed the existence of satellite lines in the neighborhood of the ground state.
These lines most likely belong to pure rotational transitions in excited vibrational states.
Under the $C_{s}$ point group, the lowest-frequency normal vibrational mode is $\nu_{9}$ ($A'$) with the anharmonic frequency predicted at 118 cm$^{-1}$.
This vibrational mode corresponds to scissoring bending vibration. As Figure~\ref{mw} shows, we were able to follow two successive excitations of this mode and to assign over 160 and 40 lines for $v_{9}=1$ and $v_{9}=2$ excited vibrational states, respectively. The assignment was carried out by comparing the values of their experimental changes in rotational constants relative to the ground state with their calculated counterparts. Only \textit{a}-type transitions have been observed for both excited vibrational states. (All measured transitions are available in a machine-readable form in the online journal.)

\begin{table*}
\caption{Vibrational changes in rotational constants for the $v_{9}$ = 1 and $v_{9}$ = 2 excited vibrational states.}
\label{excited}
\setlength{\tabcolsep}{12.0pt}
\begin{center}
\begin{tabular}{ l r r r r }
\hline\hline
  &  \multicolumn{2}{c}{$v_{9}=1$} & \multicolumn{2}{c}{$v_{9}=2$} \\
\hline
 &  Exp. & Calculated$^{a}$ & Exp. & Estimated$^{b}$ \\
$A_{v}-A_{0}$  /MHz             &  -95.334(34)   &  -102.21   &  -192.005(48)   &  -190.67      \\
$B_{v}-B_{0}$  /MHz             &   -10.4628(10)  &   -9.36  &   -20.8073(19)  &   -20.93     \\
$C_{v}-C_{0}$  /MHz             &   -1.95779(68)  &  -1.75  &   -3.8569(11)  &   -3.92    \\
\hline
\end{tabular}
\end{center}
\tablecomments{
$^{a}$Calculated at the B3PLYP/6-311++G(d,p) level.
$^{b}$Estimated from experimental changes listed for $v_{9}=1$ by assuming their additivity.
}
\end{table*}

\section{OBSERVATIONAL SEARCHES}

\paragraph{Sgr B2N} We have searched for signal from HCCSCN in the Prebiotic Interstellar Molecular Survey (PRIMOS) project observations toward Sgr B2(N) using the 100 m Robert C Byrd Green Bank Telescope (GBT).  Sgr B2(N) is a massive star-forming region \citep{Turner:1991:617} located at a distance of $\sim$8.3\,kpc \citep{Reid:2014:130}. The details of this dataset are presented elsewhere \citep{Neill:2012:153} and only a few pertinent details will be summarized.  The observations were taken in position switched mode, with the pointing position set to (J2000) $\alpha = 17^h47^m19^s.8, \delta = -28^\circ22^\prime17^{\prime\prime}$.  The bandwidth covers a total range of 1--50\,GHz with a few gaps, primarily at the lower frequencies, and at a resolution of $\sim$24\,kHz corresponding to 2--0.14\,km\,s$^{-1}$ across the band.  The RMS sensitivity ranges from a few to a few tens of mK.

\begin{figure*}[tbh!]
    \centering
    \includegraphics[width=0.49\textwidth]{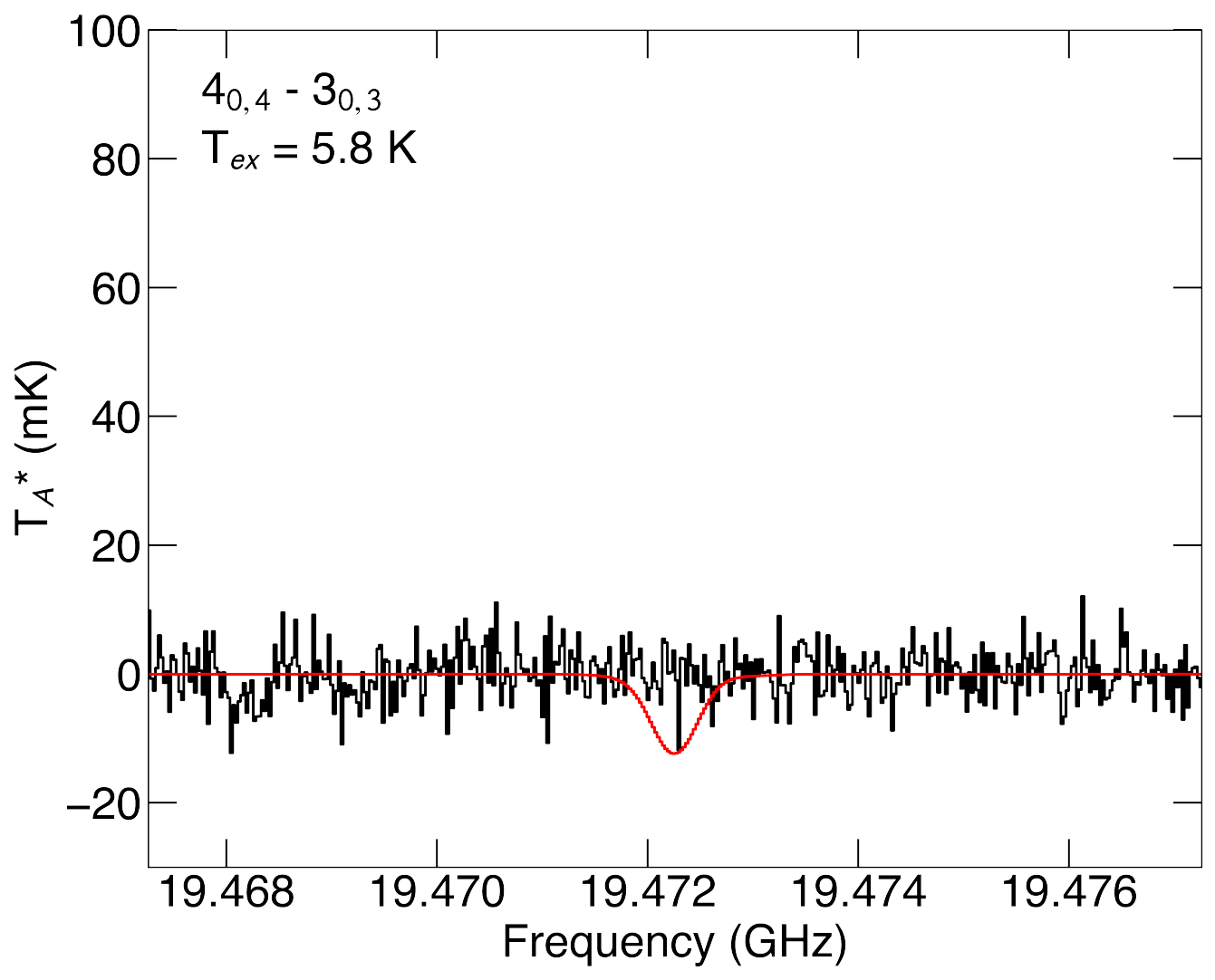}
    \includegraphics[width=0.49\textwidth]{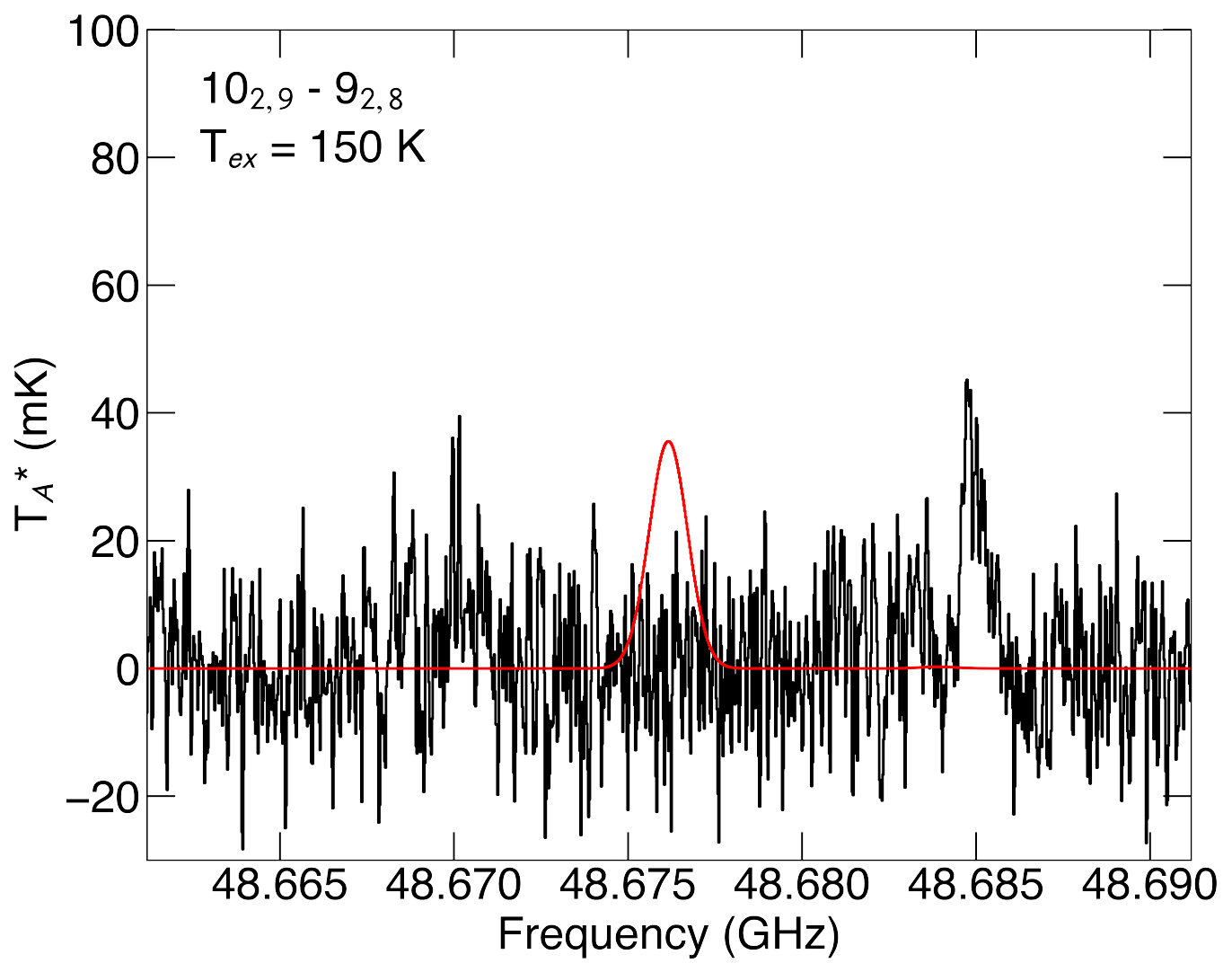}
    \caption{Transitions used to constrain the 3$\sigma$ upper limit for HCCSCN in the GBT PRIMOS observations of Sgr B2(N) in the case of a cold, extended component in absorption (left) and a warm, compact component (right).  Hyperfine splitting is not resolved at these linewidths (see text), so the unresolved quantum numbers are provided.}
    \label{primos}
\end{figure*}

Molecules in Sgr B2(N) observations typically fall into one of two categories: cold, extended, and in absorption (e.g., \citealt{McGuire:2016:1449}) or warm, compact, and in emission (e.g., \citealt{Belloche:2013:A47}).  We therefore searched for HCCSCN under two sets of conditions using a single-excitation model: a cold model with $T_{ex} = 5.8$\,K with a 20$^{\prime\prime}$ source size and a warm model with $T_{ex} = 150$\,K with a 5$^{\prime\prime}$ source size.  The linewidth was 8\,km\,s$^{-1}$ and the $V_{LSR}$ was +64\,km\,s$^{-1}$ in both cases.  At any reasonable linewidth, splitting due to hyperfine components is negligible and unresolved. The 3$\sigma$ upper limit was set using what would be the highest signal-to-noise ratio line in the case of a detection.  No signal was seen consistent with either scenario, and explorations of the data preclude any detection at other excitation temperatures as well. The 3$\sigma$ upper limit for warm HCCSCN was $N_T \leq 5\times10^{15}$\,cm$^{-2}$ and for cold HCCSCN was $N_T \leq 8\times10^{12}$\,cm$^{-2}$.  This corresponds to a fractional abundance of $X_{\ce{H2}} \leq 8\times10^{-12}$\,cm$^{-2}$ assuming an $N_{\ce{H2}} = 1\times10^{24}$\,cm$^{-2}$ \citep{Lis:1990:195}. The much more stringent limit in the case of cold HCCSCN is due to the substantially reduced partition function, which would result in far stronger transitions (relatively speaking) at these frequencies. The rotational partition function $(Q_{rot})$ was evaluated by summation of the Boltzmann factors over the energy levels in the ground vibrational state, see Table \ref{partition}. We used the SPCAT program \citep{pickett1998} to undertake this summation numerically employing the spectroscopic constants from Table \ref{constants} and all rotational states up to J= 250 and $K_{a}$= 150. Figure~\ref{primos} shows the constraining transitions in each case.

\begin{table}
\caption{Rotational and vibrational partition functions for HCCSCN at different temperatures.}
\label{partition}
\begin{center}
\setlength{\tabcolsep}{10pt}
\begin{tabular}{ r r r }
$T$ (K) &  $Q_{\text{rot}}$ & $Q_{\text{vib}}$ \\
\hline
  1000.000    & 2063131.21 & 1851.5\\
   500.000    & 724859.39 & 39.6\\
   300.000    & 335584.98 & 6.3\\ 
   225.000    & 217628.24 & 3.2 \\
   150.000    & 118273.41 & 1.7\\
    75.000    & 41753.40 & 1.1\\
    37.500    & 14755.85 & 1.0\\
    18.750    & 5219.50 & 1.0\\
     9.375    & 1848.40 & 1.0\\
     5.000    &  722.22 & 1.0\\
     3.000    & 337.21 & 1.0\\     
\hline
\end{tabular}
\end{center}
\end{table}

\paragraph{NGC 6334I} We have also searched for signal in Atacama Large Millimeter/sub-Millimeter Array (ALMA) observations of the massive star-forming region NGC 6334I.  The dataset used here is the same that was used for the first interstellar detection of methoxymethanol (\ce{CH3OCH2OH}; \citealt{McGuire:2017:L46}).  The most constraining transition is located near $\sim$280.8\,GHz in data from Project \#2015.A.00022.T.  The phase center for the observations was (J2000) $\alpha = 17^h20^m53^s.36, \delta = -35^\circ47^\prime00^{\prime\prime}.0$.  The angular resolution achieved was 0$^{\prime\prime}$.25$\times$0.$^{\prime\prime}$19, with a spectral resolution of 1.1\,km\,s$^{-1}$ and a sensitivity of 2.0\,mJy\,beam$^{-1}$.  The extraction position for the spectra that were investigated was (J2000) $\alpha = 17^h20^m53^s.373, \delta = -35^\circ46^\prime58^{\prime\prime}.14$, the same position in which \ce{CH3OCH2OH} was originally detected.  We use the same excitation temperature ($T_{ex}  = 200$\,K), linewidth ($\Delta V = 2.4$\,km\,s$^{-1}$), and $V_{lsr}$ = -7\,km\,s$^{-1}$ as in that work, and assume the source fills the beam.  As with Sgr B2(N), hyperfine splitting is unresolved.  We derive an upper limit of $N_T \leq 5\times10^{15}$\,cm$^{-2}$.  This corresponds to a fractional abundance of $X_{\ce{H2}} \leq 2\times10^{-9}$\,cm$^{-2}$ assuming an $N_{\ce{H2}} = 3\times10^{24}$\,cm$^{-2}$ \citep{Zernickel:2012:A87}.  We note that the \ce{H2} column density is derived from Herschel observations of the source and assume a homogenous gas distribution over a 10$^{\prime\prime}$ emitting region toward the hot core.  ALMA observations show this overall size is realistic, but that the underlying gas distribution is heterogenous \citep{El-Abd:2024:}.  As a result, this \ce{H2} column is a lower limit.

\begin{figure*}[tbh!]
    \centering
    \includegraphics[width=0.30\textwidth]{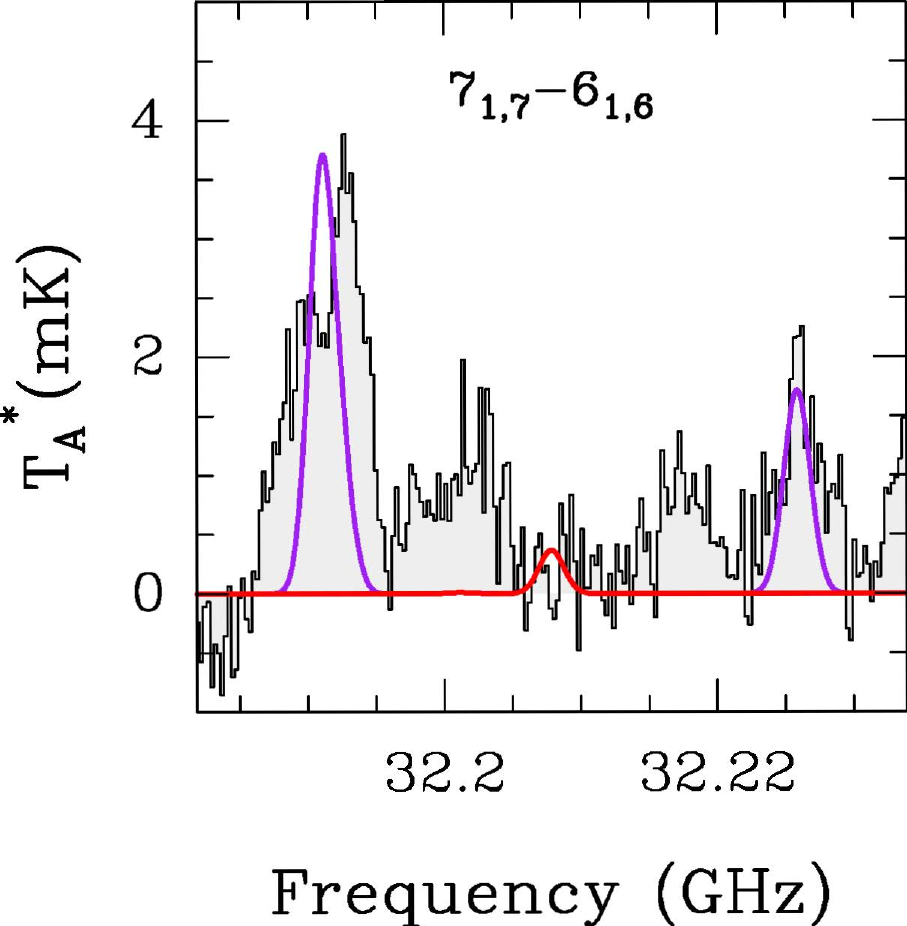}
    \hspace{1mm}
     \includegraphics[width=0.30\textwidth]{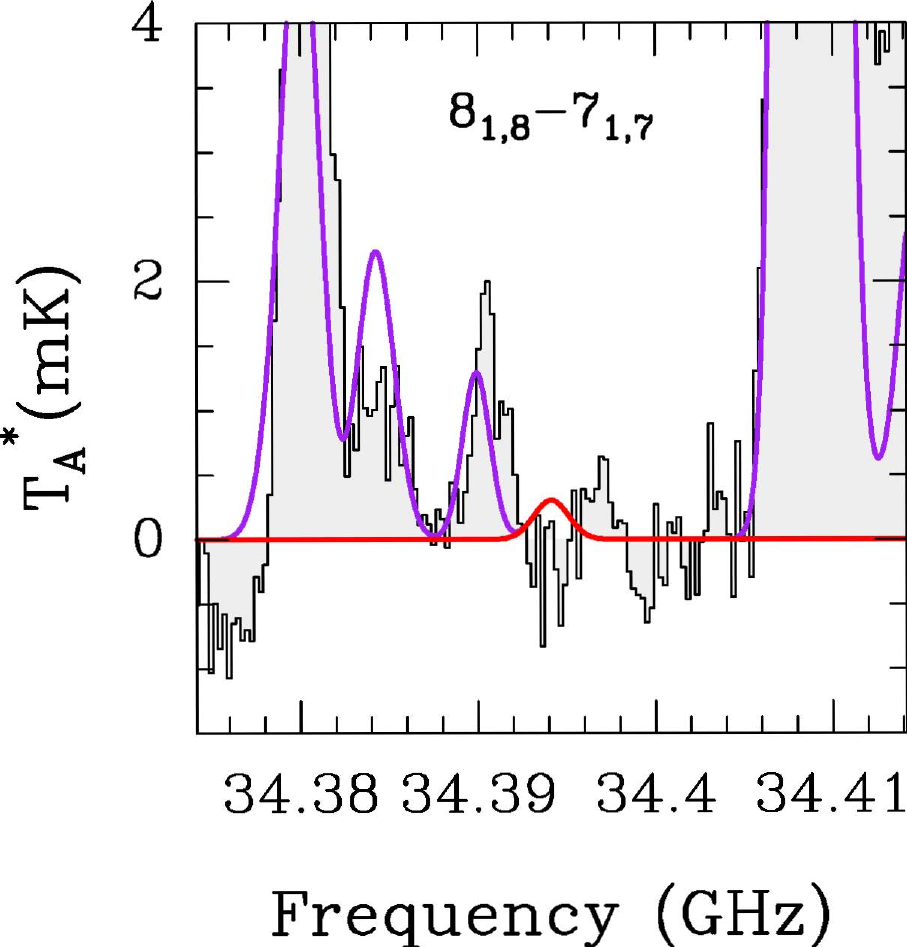}
     \hspace{1mm}
     \includegraphics[width=0.315\textwidth]{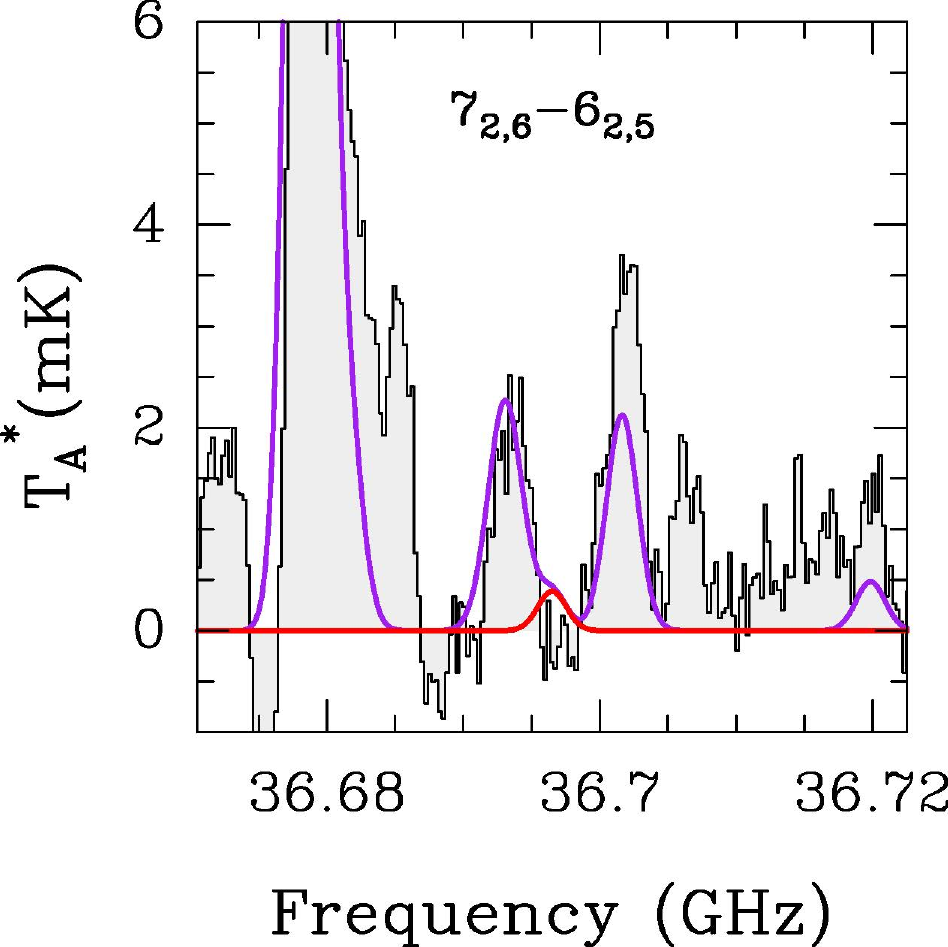}
    \caption{Transitions used to constrain the upper limit for HCCSCN towards the G+0.693-0.027 molecular cloud. The gray histogram shows the observed spectra, the red curve indicates the LTE model using the upper limit derived for the column density (see text), and the purple line shows the line emission from all the molecules identified in the region (including the contribution of HCCSCN). As in Figure \ref{primos} hyperfine splitting is not resolved, so the unresolved quantum numbers are provided.}
    \label{g0693}
\end{figure*}

\paragraph{G+0.693-0.027} We also searched for the species toward the G+0.693-0.027 (hereafter G+0.693) molecular cloud located in the Sgr B2 complex in the Galactic Center. 
This cloud is a well suited target to address the search for HCCSCN because it exhibits an extremely rich chemical content (e.g. \citealt{zeng_complex_2018,zeng2021,zeng2023,rivilla2018,rivilla_abundant_2019, rivilla2020b,rivilla2021a,rivilla2021b,rivilla2022a,rivilla2022b,rivilla2022c,rivilla2023,rodriguez-almeida2021b,jimenez-serra2022,colzi2022,
massalkhi2023,
sanz-novo2023,
san-andres2024}). In the last five years, 18 new interstellar molecules have been identified, and among them several with sulfur, such as thioformic acid (HCOSH, \citealt{rodriguez-almeida2021a}), the O-protonated carbonyl sulfide (HOCS$^{+}$, \citealt{Sanz-Novo2024a}), and thionylimide (HNSO, \citealt{Sanz-Novo2024b}). 
Moreover, it has been found that the relative abundances of S-bearing species with O-analogs are significantly higher than in other molecular clouds, suggesting that S is much less depleted in G+0.693, likely due to the action of large-scale shocks (\citealt{Sanz-Novo2024a}). 

We performed the search of HCCSCN using a high-sensitivity spectral survey carried out with the Yebes 40m and IRAM 30m telescopes.
The observations were carried out
during sessions between March 2021 and March 2022, with a total telescope scheduled time of 230\,hours, of which 110\,hours were on source.
We used the position switching mode, centering the observations at $\alpha$ = $\,$17$^{\rm h}$47$^{\rm m}$22$^{\rm s}$, $\delta$ = $\,-$28$^{\circ}$21$^{\prime}$27$^{\prime\prime}$, with the off position shifted by $\Delta\alpha$~=~$-885$$^{\prime\prime}$ and $\Delta\delta$~=~$290$$^{\prime\prime}$. 
The noise at this spectral resolution is 0.25$-$0.9 mK, depending on the spectral range.
The line intensity of the spectra was measured in units of $T_{\mathrm{A}}^{\ast}$, since the molecular emission toward G+0.693 is extended over the beams (e.g. \citealt{brunken_interstellar_2010}).
More details of these observations were provided on \citet{rivilla2023}.

We implemented the spectroscopic entry of HCCSCN presented in this work into the MADCUBA package {\footnote{Madrid Data Cube Analysis on ImageJ is a software developed at the Center of Astrobiology (CAB) in Madrid; http://cab.inta-csic.es/madcuba/}}; version 04/07/2023; \citealt{martin2019}).
Using the SLIM (Spectral Line Identification and Modeling) tool of MADCUBA, we generated a synthetic spectra under the assumption of Local Thermodynamic Equilibrium (LTE), and compared with the observed spectra.
The species is not identified in the survey. Figure \ref{g0693} shows the three brightest transitions according to the LTE model that are not blended with line emission of other molecules. We have thus derived with SLIM a 3$\sigma$ upper limit (in integrated intensity), using typical parameters derived for other species in this cloud (line width of 20 km s$^{-1}$, excitation temperature of 8 K, and a velocity of 68 km s$^{-1}$). We obtained a value of $N<$1.1$\times$10$^{12}$ cm$^{-2}$. This translates into a fractional abundance with respect to H$_2$ of $<$ 8.1 $\times$ 10$^{-12}$, using $N$(H$_{2}$) = 1.35$\times$10$^{23}$ cm$^{-2}$ from \citet{martin2008}.

\begin{figure*}
\centering
\includegraphics[scale=0.6]{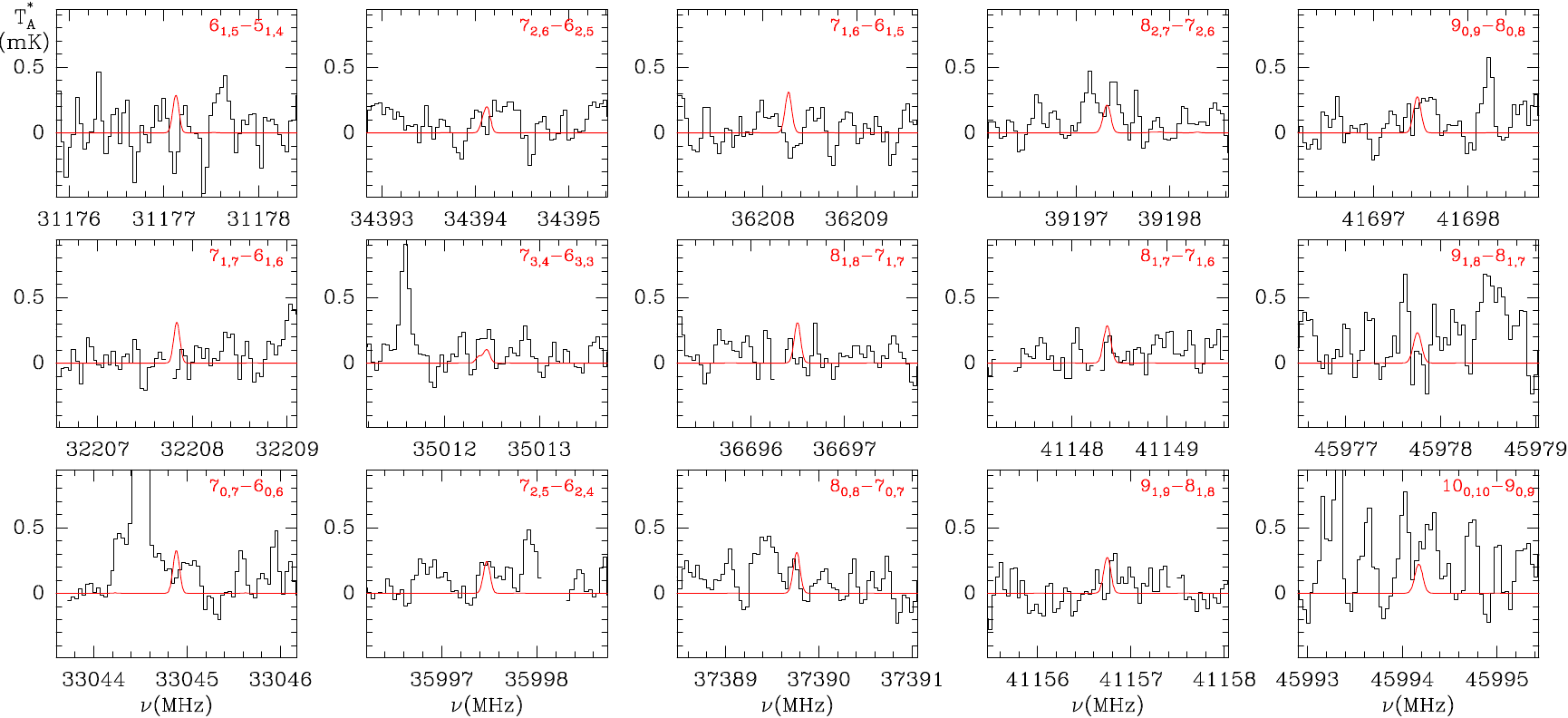}
\caption{Observed lines of HCCSCN with the QUIJOTE line survey.
The abscissa corresponds to the rest frequency in MHz.
The ordinate is the antenna temperature corrected for atmospheric and telescope losses in mK.
Blanked channels correspond to negative features produced in the folding of the frequency switching data.
The synthetic spectrum for each transition is shown in red. The physical parameters used to generate them
are given in the text.}
\label{fig_tmc1}
\end{figure*}

\paragraph{TMC-1} The observations of TMC-1 are part of the QUIJOTE\footnote{\textbf{Q}-band \textbf{U}ltrasensitive \textbf{I}nspection \textbf{J}ourney to the \textbf{O}bscure \textbf{T}MC-1 \textbf{E}nvironment} line survey \citep{Cernicharo2021d,Cernicharo2024}. The data were gathered with the Yebes 40m radio telescope equipped with the Q-band receivers and spectrometers built under the Nanocosmos project\footnote{\texttt{https://nanocosmos.iff.csic.es/}}. The selected position corresponds to the cyanopolyyne peak (CP) in TMC-1 ($\alpha_{J2000}=4^{\rm h} 41^{\rm  m} 41.9^{\rm s}$ and $\delta_{J2000}= +25^\circ 41' 27.0''$). The receivers consist of two cold high-electron mobility transistor amplifiers covering the 31.0-50.3 GHz band with horizontal and vertical polarizations.
The receiver temperatures are 16\,K at 32 GHz and 30\,K at 50 GHz. The backends are $2\times8\times2.5$ GHz  fast Fourier transform spectrometers with a spectral resolution of 38 kHz, providing the whole coverage of the Q-band in both polarizations.  A more detailed description of the system is given by \citet{Tercero2021}. 

The data of the QUIJOTE line survey presented here were gathered in several
observing runs between November 2019 and July 2023.
All observations are performed using frequency-switching observing mode with
a frequency throw of 8 and 10 MHz. The total observing time on the source
is 737 (frequency throw of 8 MHz) and 465 hours (frequency throw of 10 MHz).
Hence, the total observing time on source is 1202 hours. The
measured sensitivity varies between 0.08 mK at 32 GHz and 0.2 mK at 49.5 GHz.
The data analysis procedure has been described by \citet{Cernicharo2022}.

As a consequence of the unprecedented sensitivity of the QUIJOTE line survey, 
more than 50 new molecular species have been found in the last four years 
in TMC-1, including protonated species of abundant species, pure hydrocarbons and radicals
\citep[see, e.g.,][and references therein]{Cernicharo2024}. 
Among these new molecules it is worth noting 
the detection of
the S-bearing species NCS, HCCS, H$_2$CCS, H$_2$CCCS, C$_4$S, HCSCN, HCSCCH, HC$_4$S, HCCCS$^+$, HCCS, 
HCCS$^+$, HSO, HCNS, and NCCHCS
\citep{Cernicharo2021b,Cernicharo2021c,Cabezas2022,Fuentetaja2022,Marcelino2023,
Cernicharo2024,Cabezas_2024}. The presence of a large reservoir of S-bearing species in TMC-1, in particular of
HCSCN, HCSCCH, NCCHCS, HCNS, HCSN and HSCN \citep{Cernicharo2021c,Cernicharo2024,Cabezas_2024}, indicates that HCCSCN deserves 
to be searched in this source as it could be the CCH derivative of HSCN. The latter molecule
has been found to be a factor $\sim$3 
more abundant than HNCS with a column density of 8.3$\times$10$^{11}$ cm$^{-2}$ \citep{Cernicharo2024}.

The spectroscopic information of HCCSCN derived in this work has been implemented in the MADEX\footnote{\texttt{https://nanocosmos.iff.csic.es/?page$\_$id=1619}} code. Because of the much larger value of the $a$-component of the dipole moment, only $a$-type transitions have been searched for. As the kinetic temperature of  the cloud is 9\,K \citep{Agundez2023}, we limited the energy of the upper 
levels of the transitions to 14\,K ($J$=10, $K_a$=0,1). In order to model the expected intensity we have assumed a source of uniform brightness temperature
with a diameter of 80$''$ \citep{Fosse2001,Cernicharo2023}. The adopted linewidth is 0.6 km\,s$^{-1}$, i.e., identical to that derived for similar species such as the CCH and CN derivative of thioformaldeyde, HCSCCH and HCSCN \citep{Cernicharo2021c}, and the CN derivative of thioketene, NCCHCS \citep{Cabezas_2024}. We also adopted a rotational temperature of 5\,K similar to that derived for HCSCN. A total of fifteen individual transitions were searched as shown in Figure \ref{fig_tmc1}. Only a few of the generated synthetic spectra have a counterpart in the data and, hence, we conclude that the molecule is not present in TMC-1 with a 3$\sigma$ upper limit to its column density, based on each individual line, of 5$\times$10$^{10}$cm$^{-2}$.

\section{ASTROCHEMICAL IMPLICATIONS} \label{sec:astrochemical}


    Sulfur molecules, in general, are of great astrochemical interest, in part, because of the longstanding mystery of ``missing sulfur,'' based on the apparent drop in observable sulfur species when comparing diffuse environments - where the abundances are roughly the solar values - to dense molecular clouds - where the combined abundances of all sulfur-bearing species can be three orders of magnitude less. It seems fairly clear that the sulfur is not disapearing, per se, but rather, is being chemically converted to some reservoir species, which has not yet been conclusively identified. Sulfur is also an important element for Terrestrial organisms, being a key constituent of biomolecules such as the amino acids cysteine and methionine, and thus, a better understanding of interstellar sulfur chemistry is an important step in understanding the inventory of prebiotic molecules which may be available for delivery to the surfaces of exoplanets and implicated in the origins of life. 

    In astrophysical environments, the synthesis pathways used to produce HCCSCN for this study would not occur. Indeed, unsaturated species such as HCCSCN are typical of the kinds of molecules formed via gas-phase reactions, which are almost exclusively barrierless, exothermic, two-body processes given the low temperatures ($\sim$ 10 K) and densities ($\sim\;10^{4}$ cm$^{-3}$) of molecular clouds. To the best of our knowledge, the only such reaction leading to HCCSCN that has been the subject of laboratory investigations is

    \begin{equation}
        \ce{HCCH + SCN -> HCCSCN + H}
    \end{equation}

    \noindent
    though, as reported by \cite{Chen2003}, this is an endothermic process with a substantial barrier, which is consistent with the upper limit of 10$^{-14}$ cm$^3$ s$^{-1}$ experimentally determined for the rate coefficient by \cite{Baren1999} in the temperature range 298-620 K. Therefore, this is not a viable formation pathway in cold astrophysical environments. Nevertheless, recent observations of other sulfur-bearing species in TMC-1 \citep{Cernicharo2021b,Cernicharo2021c,Cernicharo2024} yield some clues that could point to other possible formation pathways to HCCSCN. One possibility are the reactions of CN with \ce{H2CCS} and \ce{HCCSH}. The first one is calculated to be endothermic by 10.3 kcal mol$^{-1}$, although the second is exothermic and could provide a valid route. Another possibility are the reactions of HCCS with HCN and HNC, although in this case the two are calculated to be endothermic by 44.1 and 29.2 kcal mol$^{-1}$, respectively. The exploration of the potential energy surface for the different reactions was done using the Molpro 2020.2 \textit{ab initio} program package \citep{MOLPRO_brief}. All the stationary points were fully optimized using the coupled cluster method (CCSD(T)) \citep{RAGHAVACHARI1989479} along with Dunning’s correlation consistent polarized valence triple-z (cc-pVTZ) basis set \citep{dunningCCST}, and the energies are relative to that of the separated reactants. The reaction \ce{C2S + H2CN} is calculated to be exothermic although it is likely that it occurs with a barrier because there is an important rearrangement. Finally, the reaction \ce{C2H + HSCN} could be a source of HCCSCN at low temperatures because it is calculated to be exothermic and could occur with a minimal rearrangement.

    However formed, gas-phase \ce{HCCSCN} could be depleted via, e.g., photodissociation, reactions with other gas-phase species, or accretion onto grains. Once accreted onto the surfaces of dust grains or dust-grain ice mantles, the dominant chemical destruction pathway is via reaction with hydrogen atoms, which can remain comparatively mobile even at the very low temperature of 10 K. Previous quantum-chemical calculations of the reaction of highly unsaturated interstellar molecules with H has found that such processes are often barrierless and exothermic, and yield one or more stable, more highly saturated end products \citep{shingledecker_case_2019,shingledecker_grain-surface_2022}. If formed on the surfaces of grains in cold molecular clouds, these stable, more highly-saturated products can become covered by the continuous accretion of yet more gas-phase species, thereby becoming trapped within the dust-grain ice mantle. If the astrophysical environment subsequently warms, such as happens during star-formation, these trapped molecules can be liberated into the gas-phase, where they can be observed. Thus, it is often possible to surmise a chemical link between an unsaturated carbon chain molecule seen in a cold core with a more saturated species seen in a hot core. In the absence of detailed quantum calculations, it is not possible to say with a high degree of certainty what the grain-surface saturation of \ce{HCCSCN} would yield, but we here suggest two possibilities by way of speculation. The first possibility is the species \ce{C3H5NS}, i.e. the ethyl ester of thiocyanic acid. Another, more astrochemically interesting possible saturation product is \ce{C3H3NS}, the heterocyclic molecule thiazole. This latter species would be of particular interest since, though now several cyclic or polycyclic species have been detected in the ISM, there are to date no detections of heterocycles with a sulfur atom, and certainly not of cycles with two heteroatoms such as thiazole, and if it were detected, it would suggest a grain-surface origin for other similar species. 

\section{CONCLUSIONS}    
We have determined for the first time precise rotational and centrifugal distortion constants for the \ce{CCH} derivative of \ce{HSCN} (\ce{HCCSCN}) from accurate laboratory measurements. This new laboratory data allowed us to search for it towards several interstellar sources without success. Several possible formation paths of this species are discussed, emphasizing that it could be formed from the reaction of \ce{CCH} with \ce{HSCN}. Similar reactions have been previously proposed for other S-bearing species found in TMC-1 such as
\ce{HCSCN}, \ce{HCSCCH}, and \ce{NCCHCS} which are \ce{CN} and \ce{CCH} derivatives of \ce{H2CS} and  \ce{H2CCS} are very abundant in that cloud. Although no spectral features of \ce{HCCSCN} were found, present data can be used with confidence in future searches in other interstellar sources.

\begin{acknowledgments}
The National Radio Astronomy Observatory is a facility of the National Science Foundation operated under cooperative agreement by Associated Universities, Inc. This paper makes use of the following ALMA data: ADS/JAO.ALMA\#2015.A.00022.T. ALMA is a partnership of ESO (representing its member states), NSF (USA) and NINS (Japan), together with NRC (Canada), MOST and ASIAA (Taiwan), and KASI (Republic of Korea), in cooperation with the Republic of Chile. The Joint ALMA Observatory operated by ESO, AUI/NRAO and NAOJ. E.R.A. and I.L. acknowledge funding from Ministerio de Ciencia e Innovación (PID2019-111396GB-I00), and Junta de Castilla y León (VA244P20). V.M.R., I.J-S. and J.M-P acknowledges support from the grant No. PID2022-136814NB-I00 by the Spanish Ministry of Science, Innovation and Universities/State Agency of Research MICIU/AEI/10.13039/501100011033 and by ERDF, UE.
VMR also acknowledges support from the grant number RYC2020-029387-I funded by MICIU/AEI/10.13039/501100011033 and by "ESF, Investing in your future", and from the Consejo Superior de Investigaciones Cient{\'i}ficas (CSIC) and the Centro de Astrobiolog{\'i}a (CAB) through the project 20225AT015 (Proyectos intramurales especiales del CSIC). J.C., C.C., and M.A. thank the Spanish Ministry of Science, Innovation and Universities for funding support through projects PID2019-106110GB-I00, and PID2019-106235GB-I00 and ERC for funding through grant ERC-2013-Syg-610256-NANOCOSMOS. M.A. thanks the Consejo Superior de Investigaciones Cient\'ificas (CSIC; Spain) for funding through project PIE 202250I097.J.-C.G. thanks the Programme National “Physique et Chimie du Milieu Interstellaire” (PCMI) of CNRS Terre @ Univers with CNRS Physique \& CNRS Chimie, co-funded by CEA and CNES.

\end{acknowledgments}


\bibliography{library,shingledecker,brett,rivilla}{}
\bibliographystyle{aasjournal}



\end{document}